\documentclass[nohyper]{JHEP3}
\usepackage{cite,epsfig,amssymb,amsmath}

\newcommand{\mycomm}[1]{\hfill\break $\phantom{a}$\kern-3.5em{\tt===$>$ \bf #1}\hfill\break}
\newcommand{\mycommA}[1]{\hfill\break $\phantom{a}$\kern-3.5em{\tt   $>$ \bf #1}\hfill\break}

\newcommand{\be}{\begin{equation}}
\newcommand{\ee}{\end{equation}}
\newcommand{\ba}{\begin{eqnarray}}
\newcommand{\ea}{\end{eqnarray}}

\catcode`\@=11 
\def\lsim{\mathrel{\mathpalette\@versim<}}
\def\gsim{\mathrel{\mathpalette\@versim>}}
\def\@versim#1#2{\vcenter{\offinterlineskip
        \ialign{$\m@th#1\hfil##\hfil$\crcr#2\crcr\sim\crcr } }}
\catcode`\@=12 

\title{Large-$x$ structure of  physical evolution kernels in Deep Inelastic Scattering}

\author{G. Grunberg\\
        Centre de Physique Th\'eorique,  \'Ecole
Polytechnique\\
        91128 Palaiseau Cedex, France\\
        E-mail: \email{georges.grunberg@cpht.polytechnique.fr}}

\abstract{The modified evolution equation for parton distributions of Dokshitzer, Marchesini and Salam is extended to non-singlet Deep Inelastic Scattering coefficient functions and the physical evolution kernels which govern their scaling violation. Considering the $x\rightarrow 1$ limit, it is found that the leading  next-to-eikonal   logarithmic contributions to the physical kernels at any loop order can be expressed in term of the one-loop cusp anomalous dimension, a result which can  presumably be extended to all orders in $(1-x)$, and  has eluded so far  threshold resummation. Similar results are shown to hold for fragmentation functions in semi-inclusive $e^{+}e^{-}$ annihilation. Gribov-Lipatov relation is found to be satisfied by the leading logarithmic part of the modified physical evolution kernels.
}

\preprint{ }

\begin{document}

\section{Introduction}

There has been recently renewed interest \cite{Kramer:1996iq,Grunberg:2007nc,Laenen:2008ux,Laenen:2008gt,Moch:2009mu,Grunberg:2009yi,Moch:2009hr,Grunberg:2009am,Soar:2009yh} 
in threshold resummation of ``next-to-eikonal'' logarithmically enhanced terms which are  suppressed by some power of the gluon energy $(1-x)$ for $x\rightarrow 1$ in momentum space (or by some power of $1/N$, $N\rightarrow\infty$ in moment space). In particular, in 
\cite{Grunberg:2007nc,Moch:2009mu,Grunberg:2009yi,Moch:2009hr,Grunberg:2009am,Soar:2009yh}
this question has been investigated at the level of ``physical evolution kernels''  which control the scaling violation of (non-singlet) structure functions.
The scale--dependence of the  Deep Inelastic Scattering (DIS) coefficient function
${\cal C}_2(x, Q^2,\mu^2_F)$ corresponding to the flavor  non-singlet $F_2(x,Q^2)$ structure function($F_2(x,Q^2)/x={\cal C}_2(x, Q^2,\mu^2_F)\otimes q_{2,ns}(x,\mu^2_F)$, where $ q_{2,ns}(x,\mu^2_F)$ is the corresponding quark distribution)
 can be expressed in terms of ${\cal C}_2(x, Q^2,\mu^2_F)$ itself, yielding the following ``physical'' evolution equation
(see e.g.~Refs.~\cite{Furmanski:1981cw,Grunberg:1982fw,Catani:1996sc,Blumlein:2000wh,vanNeerven:2001pe}):
\begin{equation}
\label{F_2_evolution}
\frac{\partial{\cal C}_2(x,Q^2,\mu^2_F)}{\partial\ln Q^2}\,=\,\int_x^1 \frac{dz}{z}\,K(z,a_s(Q^2))\,{\cal C}_2(x/z,Q^2,\mu^2_F)\equiv K(x,a_s(Q^2))\otimes{\cal C}_2(x,Q^2,\mu^2_F)\ ,
\end{equation}
where $\mu_F$ is the factorization scale (I assume for definitness the $\overline{MS}$ factorization scheme is used).
$K(x,a_s(Q^2))$ is the momentum space \emph{physical evolution kernel}, or {\em physical anomalous dimension}; it is independent of the factorization scale and renormalization--scheme invariant.

 In \cite{Gardi:2007ma},  the result for the leading  contribution to this quantity in the $x\rightarrow 1$ limit  was derived, which resums all  logarithms at the leading eikonal level, and nicely summarizes analytically in {\em momentum} space the standard results \cite{Sterman:1986aj,Catani:1989ne} of threshold resummation:
\begin{equation}
\label{K_x}
K(x,a_s(Q^2))\sim \frac{{\cal J}\left(rQ^2\right)}{r}\,+B_{\delta}^{DIS}(a_s(Q^2))\,\delta(1-x)\ , \end{equation}
where $r=\frac{1-x}{x}$ (with $rQ^2\equiv W^2$  the final state  ``jet'' mass),
$B_{\delta}^{DIS}(a_s)$ 
 is related to the the quark form factor, and ${\cal J}(Q^2)$, the ``physical Sudakov anomalous dimension'' (a renormalization scheme invariant quantity), is given by:

\begin{eqnarray} \label{standard-J-coupling}
{\cal J}(Q^2)&= &A\left(a_s(Q^2)\right)+{dB\left(a_s(Q^2)\right)\over d\ln Q^2}\\
&=&A\left(a_s(Q^2)\right)+\beta\left(a_s(Q^2)\right)\frac{dB\left(a_s(Q^2)\right)}{da_s}\equiv\sum_{i=1}^\infty
j_i  a_s^{i}(Q^2)\nonumber
\ .\end{eqnarray}
In eq.(\ref{standard-J-coupling}),
\begin{equation}\label{cusp}
A(a_s)=\sum_{i=1}^\infty
A_ia_s^{i}
\end{equation} 
 is the universal ``cusp'' anomalous dimension \cite{Korchemsky:1993uz} (see also \cite{Catani:1990rr}), with $a_s\equiv\frac{\alpha_s}{4\pi}$  the $\overline{MS}$ coupling, 
 
 \begin{equation}\beta(a_s)=\frac{d a_s}{d\ln Q^2}=-\beta_0\, a_s^2-\beta_1\, a_s^3-\beta_2\, a_s^4+...\label{beta}\end{equation}
is the beta function (with $\beta_0=\frac{11}{3}C_A-\frac{2}{3}n_f$)
and
\begin{equation}\label{B-stan}
B(a_s)=
\sum_{i=1}^\infty B_i a_s^{i}
\end{equation}
is the usual final state ``jet function''  anomalous dimension.  It should be noted that $j_1=A_1$ (the one loop cusp anomalous dimension), and also that both $A(a_s)$ and $B(a_s)$ (in contrast to ${\cal J}(Q^2)$) are
renormalization scheme-dependent quantities.
The renormalization group invariance of ${\cal J}(Q^2)$ yields the standard relation:

\begin{eqnarray}{\cal J}\left((1-x)Q^2\right)&=&j_{1}\ a_s+a_s^2[-j_{1}\beta_0 L_x+j_{2}]\label{j-expand}\\
&+ &a_s^3[j_{1}\beta_0^2 L_x^2-(j_{1}\beta_1+2 j_{2} \beta_0) L_x+j_{3}]\nonumber\\
&+ &a_s^4[-j_{1}\beta_0^3 L_x^3+(\frac{5}{2}j_{1}\beta_1\beta_0+3 j_{2} \beta_0^2) L_x^2-(j_{1}\beta_2+2 j_{2} \beta_1+3 j_{3} \beta_0) L_x+j_{4}]+...\ ,\nonumber
\end{eqnarray}
where  $L_x\equiv \ln(1-x)$ and $a_s=a_s(Q^2)$, from which the structure of all the eikonal logarithms in $K(x,a_s(Q^2))$ can be derived. A term like $\frac{L_x^p}{1-x}$ arising from $ \frac{{\cal J}\left(rQ^2\right)}{r}$ in eq.(\ref{K_x}) must be interpreted as usual as a standard $+$-distribution. {\em All} the eikonal logarithms are thus absorbed into the {\em single} scale $(1-x)Q^2$ (see also \cite{Amati:1980ch,Ciafaloni:1980pz,Ciafaloni:1981nm} and section VI-E in \cite{Grunberg:1982fw}).

However, no analogous result holds \cite{Grunberg:2009yi} at the next-to-eikonal level (except \cite{Grunberg:2007nc} at large-$\beta_0$). In this note, I show that the {\em leading} 
next-to-eikonal logarithmic contributions to the physical evolution kernel at a given order in  $a_s$  can actually be determined in term of lower order  {\em leading} eikonal coefficients, representing the first step towards threshold resummation at the next-to-eikonal level. This result is obtained by extending the approach of \cite{Dokshitzer:1995ev,Dokshitzer:2005bf} (which deals with parton distributions) to the DIS coefficient functions themselves.

\section{The modified physical kernel}

I consider the class of modified physical evolution equations:

\begin{equation}
\label{F_2_new-evolution}
\frac{\partial {\cal C}_2(x,Q^2,\mu^2_F)}{\partial \ln Q^2}\,=\,\int_x^1 \frac{dz}{z}\,K(z,a_s(Q^2),\lambda)\,{\cal C}_2(x/z,Q^2/z^{\lambda},\mu^2_F)\ ,
\end{equation}
where for book-keeping purposes I introduced the parameter $\lambda$, which shall eventually be set to its physically meaningful value  $\lambda=1$, in straightforward analogy to the modified evolution equation for parton distributions of \cite{Dokshitzer:2005bf}. I note that $K(x,a_s,\lambda=0)\equiv K(x,a_s)$, the `standard' physical evolution kernel. Eq.(\ref{F_2_new-evolution}) allows to determine $K(x,a_s,\lambda)$ given $ K(x,a_s)$ (or vice-versa). Indeed, expanding ${\cal C}_2(y,Q^2/z^{\lambda},\mu^2_F)$ around $z=1$, keeping the other two variables fixed, and reporting into eq.(\ref{F_2_new-evolution}), one easily derives the following relation between $K(x,a_s,\lambda)$ and $ K(x,a_s)$: 

\begin{eqnarray}
\label{new-kernel-full}K(x,a_s)&=&K(x,a_s,\lambda)-\lambda [\ln x\  K(x,a_s,\lambda)]\otimes K(x,a_s)\\
& &+\frac{\lambda^2}{2}[\ln^2 x\ K(x,a_s,\lambda)]\otimes[\beta(a_s)\frac{\partial K(x,a_s)}{\partial a_s}+K(x,a_s)\otimes K(x,a_s)] +...\nonumber\ , \end{eqnarray}
where only terms with a single overall factor of $\lambda$ need actually to be kept up to next-to-eikonal order, since one can check terms with more  factors of $\lambda$, which are associated to more factors of $\ln x$,  are not relevant  to determine the  next-to-eikonal logarithms in the physical kernel. In the rest of the paper (except section 4) I shall therefore simply use:

\begin{equation}
\label{new-kernel}K(x,a_s)=K(x,a_s,\lambda)-\lambda [\ln x\  K(x,a_s,\lambda)]\otimes K(x,a_s)+...\ . \end{equation}
 Eq.(\ref{new-kernel}) can be solved perturbatively. Setting:

\begin{equation}
\label{new-kernel-pert}K(x,a_s,\lambda)=K_0(x,\lambda)a_s+K_1(x,\lambda)a_s^2+K_2(x,\lambda)a_s^3+K_3(x,\lambda)a_s^3+...\end{equation}
(and similarly for $K(x,a_s)$), one gets:

\begin{eqnarray}
\label{new-kernel-bis}
K_0(x,\lambda)&=&K_0(x)\\
K_1(x,\lambda)&=&K_1(x)+\lambda [\ln x\  K_0(x)] \otimes K_0(x)\nonumber\\
K_2(x,\lambda)&=&K_2(x)+\lambda \{[\ln x\  K_1(x)]\otimes K_0(x)+[\ln x\  K_0(x)]\otimes K_1(x)\}+...\nonumber\\
K_3(x,\lambda)&=&K_3(x)+\lambda \{[\ln x\  K_2(x)]\otimes K_0(x)+[\ln x\  K_1(x)]\otimes K_1(x)\nonumber\\
& &+[\ln x\  K_0(x)]\otimes K_2(x)\}+...\nonumber
\ . \end{eqnarray}
The $K_i(x)$'s are determined in term of splitting functions and coefficient functions as follows \cite{vanNeerven:2001pe}:

\begin{eqnarray}\label{Ki} K_0(x)&=&P_0(x)\\
K_1(x)&=&P_1(x)-\beta_0\   c_1(x)\nonumber\\
K_2(x)&= &P_2(x)-\beta_1\  c_1(x)-\beta_0(2 c_2(x)-c_1^{\otimes 2}(x))\nonumber\\
K_3(x)&= &P_3(x)-\beta_2\  c_1(x)-\beta_1(2 c_2(x)-c_1^{\otimes 2}(x))-\beta_0(3 c_3(x)-3c_2(x)\otimes c_1(x)+c_1^{\otimes 3}(x))\nonumber\ , \end{eqnarray}
where $P_i(x)$ are the standard $(i+1)$-loop splitting functions,  $c_i(x)$ are the $i$-loop coefficient functions, and $c_1^{\otimes 2}(x)\equiv c_1(x)\otimes c_1(x)$, etc....

Consider now the $x\rightarrow 1$ limit. The one-loop splitting function is given by \cite{Altarelli:1977zs}:

\begin{equation}\label{P0}P_0(x)=A_1p_{qq}(x)+B_1^{\delta}\, \delta(1-x)\ ,\end{equation} 
with $A_1=4C_F$,   and\footnote{$p_{qq}(x)$ is defined to be $1/2$ the corresponding function in \cite{Moch:2009hr}.}: 

\begin{equation}
\label{prefactor}p_{qq}(x)=\frac{1}{1-x}-1+\frac{1}{2}(1-x)=\frac{x}{1-x}+\frac{1}{2}(1-x)=\frac{1}{r}+\frac{1}{2}(1-x)\ .
 \end{equation}
 Moreover, at the next-to-eikonal level we have, dropping from now on $\delta$ function  contributions:

\begin{equation}
\label{P1}
P_1(x)=\frac{A_2}{r}+C_2L_x+D_2+...\ ,
\end{equation}
with \cite{Curci:1980uw}:
  
\begin{equation}\label{C2}C_2=A_1^2\ .\end{equation}  
 Also:

\begin{equation}
\label{c1}
c_1(x)=\frac{c_{11}L_x+c_{10}}{r}+b_{11}L_x+b_{10}+...
\end{equation}
with $c_{11}=A_1=4C_F$,  $b_{11}=0$.
From eq.(\ref{Ki}) one can derive \cite{Grunberg:2009yi,Moch:2009hr}  the following expansions for $x\rightarrow 1$:

\begin{eqnarray}
\label{large-x-kernel}
K_0(x)&=&P_0(x)=\frac{k_{10}}{r}+h_{10}+...\\
K_1(x)&=&\frac{k_{21}L_x+k_{20}}{r}+h_{21}L_x+h_{20}+...\nonumber\\
K_2(x)&=&\frac{k_{32}L_x^2+k_{31}L_x+k_{30}}{r}+h_{32}L_x^2+h_{31}L_x+h_{30}+...\nonumber\\
K_3(x)&=&\frac{k_{43}L_x^3+k_{42}L_x^2+k_{41}L_x+k_{40}}{r}+h_{43}L_x^3+h_{42}L_x^2+h_{41}L_x+h_{40}+...\nonumber\ . \end{eqnarray}

\section{Leading next-to-eikonal logarithms}

\subsection{Two loop kernel}
From eq.(\ref{Ki}), (\ref{P0}), (\ref{P1}) and (\ref{c1}) one deduces:
$k_{10}=A_1$, $h_{10}=0$,  and $k_{21}=-\beta_0 A_1$, $h_{21}=C_2$. Then eq.(\ref{new-kernel-bis}) yields for $x\rightarrow 1$:

\begin{eqnarray}
\label{large-x-new-kernel-1}
K_0(x,\lambda)&=&P_0(x)\\
K_1(x,\lambda)&=&\frac{k_{21}L_x+k_{20}}{r}+(h_{21}-\lambda\  k_{10}^2)L_x+{\cal O}(L_x^0)\nonumber\ . \end{eqnarray}
Now

\begin{equation}
\label{next-to-eik-LL-1}
h_{21}(\lambda)=h_{21}-\lambda\  k_{10}^2=C_2-\lambda\  A_1^2=(1-\lambda)A_1^2\ .\end{equation}
Thus, setting  $\lambda=1$, one  finds that the leading  next-to-eikonal  logarithm in $K_1(x,\lambda=1)$ vanishes, yielding the relation:

\begin{equation}
\label{next-to-eik-LL-1-bis}
h_{21}=k_{10}^2=16 C_F^2\  , \end{equation}
which is correct \cite{Grunberg:2009yi,Moch:2009hr}.
This finding is not surprising: up to two loop,  the leading  next-to-eikonal  logarithm is contributed only by the  splitting function, since $b_{11}=0$   (e.g. $h_{21}=C_2$), and one effectively recovers the result (eq.(\ref{C2})) holding \cite{Dokshitzer:2005bf}
 for the two loop splitting function. The situation however changes drastically at three loop, where the leading  next-to-eikonal  logarithm is contributed by the coefficient function rather then the splitting function, and the crucial question is whether the leading next-to-eikonal  logarithm still vanishes for $\lambda=1$.

\subsection{Three loop kernel}
 Eq.(\ref{new-kernel-bis}) yields for $x\rightarrow 1$:

\begin{equation}
\label{large-x-new-kernel-2}
K_2(x,\lambda)=\frac{k_{32}L_x^2+k_{31}L_x+k_{30}}{r}+(h_{32}-\lambda\frac{3}{2}k_{21}k_{10})L_x^2+{\cal O}(L_x)
\ . \end{equation}
Requiring $h_{32}(\lambda)$,   the coefficient of the ${\cal O}(L_x^2)$ term, to vanish for $\lambda=1$ predicts:

\begin{equation}
\label{prediction1}h_{32}=\frac{3}{2}k_{21}k_{10}=-\frac{3}{2}\beta_0 A_1^2=-24 \beta_0 C_F^2\ , \end{equation}
which is indeed the correct \cite{Grunberg:2009yi,Moch:2009hr} value. I stress that this result is {\em not} a consequence of the  relation \cite{Moch:2004pa,Dokshitzer:2005bf,Basso:2006nk} $C_3=2A_1A_2$ for $P_2(x)$. Indeed it is well-known \cite{Korchemsky:1988si} that the  $P_i(x)$'s, and in particular $P_2(x)$, have only a {\em single} next-to-eikonal logarithm:

\begin{equation}
\label{P2}
P_2(x)= \frac{A_3}{r}+C_3L_x+D_3+...\  ,\end{equation}
and thus $P_2(x)$ cannot contribute to the {\em double} logarithm in $K_2(x)$. Rather, $h_{32}$ is contributed  by the coefficient functions in eq.(\ref{Ki}), 
and eq.(\ref{prediction1}) yields a prediction for the ${\cal O}(L_x^2)$ term in $c_2(x)$.

\subsection{Four loop kernel}

Eq.(\ref{new-kernel-bis}) yields for $x\rightarrow 1$:

\begin{eqnarray}
\label{large-x-new-kernel-3}
K_3(x,\lambda)&=&\frac{k_{43}L_x^3+k_{42}L_x^2+k_{41}L_x+k_{40}}{r}\nonumber\\
& &+[h_{43}-\lambda (\frac{4}{3} k_{10}k_{32}+\frac{1}{2}k_{21}^2)]L_x^3+{\cal O}(L_x^2)\ , \end{eqnarray}
where $k_{32}=A_1\beta_0^2$ (consistently with eq.(\ref{j-expand})).
Requiring $h_{43}(\lambda)$, the coefficient of the ${\cal O}(L_x^3)$ term, to vanish for $\lambda=1$ predicts:

\begin{equation}
\label{prediction2}h_{43}=\frac{4}{3} k_{10}k_{32}+\frac{1}{2}k_{21}^2=\frac{11}{6}\beta_0^2 A_1^2=\frac{88}{3} \beta_0^2 C_F^2\ , \end{equation}
which is again the correct \cite{Grunberg:2009yi,Moch:2009hr} value.

\subsection{Five loop kernel}
One can similarly predict  the leading next-to-eikonal logarithm in the five loop physical kernel (which depends on the four loop coefficient function). Using eq.(\ref{new-kernel}), the coefficient of the ${\cal O}(L_x^4)$ term in $K_4(x,\lambda)$ is found to be given by:

\begin{equation}
\label{5loop-log}h_{54}(\lambda)=h_{54}-\lambda (\frac{5}{4} k_{10}k_{43}+\frac{5}{6}k_{21}k_{32})\ , \end{equation}
where $k_{43}=-A_1\beta_0^3$ (again consistent with eq.(\ref{j-expand})).
Requiring this coefficient to vanish for $\lambda=1$ predicts\footnote{Hence $\xi_{\rm DIS_4}=\frac{100}{3}$ in the notation of \cite{Moch:2009hr}.}:

\begin{equation}
\label{prediction3}h_{54}=\frac{5}{4} k_{10}k_{43}+\frac{5}{6}k_{21}k_{32}=-\frac{25}{12}\beta_0^3 A_1^2=-\frac{100}{3} \beta_0^3 C_F^2\ . \end{equation}

\subsection{All-order relations}
Defining moments by
\begin{equation}
\label{moments}K(N,a_s)=\int_0^1 dx\, x^{N-1} K(x,a_s)\ , \end{equation}
eq.(\ref{new-kernel}) yields in moment space:

\begin{equation}
\label{new-kernel-moments}K(N,a_s)=\frac{K(N,a_s,\lambda)}{1+\lambda\dot {K}(N,a_s,\lambda)}\ , \end{equation}
where $\dot{f}\equiv \partial f/\partial N$. Assuming the leading next-to-eikonal  logarithms vanish to all orders (in an expansion in $1/r$) for $\lambda=1$, i.e. that $h_{i+1,i}(\lambda=1)=0$ for $i\geq 0$, one can derive \cite{GG} from eq.(\ref{new-kernel-moments}) the resummation formula:

\begin{equation}
\label{all-order-LL}
\sum_{i=0}^\infty h_{i+1,i} L_x^i a_s^{i+1}=\frac{A_1}{\beta_0}\frac{A_1 a_s}{1+a_s \beta_0 L_x}\ln(1+a_s \beta_0 L_x)\ ,
 \end{equation}
which correctly reproduces the results in the previous subsections. 

\noindent One can further show \cite{GG} that the  moment space functional relation which accounts for {\em leading} logarithms at all orders in $(1-x)$ is:

\begin{equation}
\label{functional-LL}K(N,a_s)=K[N-\lambda K(N,a_s),a_s,\lambda]\ .
 \end{equation}
  Eq.(\ref{new-kernel-moments}) results from expanding the right hand side of eq.(\ref{functional-LL}) to  first order in $\Delta N\equiv\lambda K(N,a_s)$. It is interesting that eq.(\ref{functional-LL}) is identical to the functional relation\footnote{A similar functional relation has been obtained in a different context in \cite{Mueller:1983js}.} obtained \cite{Basso:2006nk,Dokshitzer:2006nm} for the splitting functions in the conformal limit (where the splitting functions coincide with the $K_i$'s).

\section{Leading next-to-next-to-eikonal logarithms}
It can be checked \cite{GG} that similar methods allow to predict using eq.(\ref{new-kernel-full}) the {\em leading} logarithmic contributions at the next-to-next-to-eikonal level, i.e. the coefficient of the $(1-x)L_x^i$ term in $K_i(x)$. The crucial new point, however, is that the leading term in the eikonal expansion has to be defined in term of the one-loop splitting function prefactor $p_{qq}(x)$ (eq.(\ref{prefactor})),
 instead of $1/r$ as in eq.(\ref{large-x-kernel}). Namely, keeping only leading logarithms at each eikonal order, the predicted $f_{j i}^c$ coefficients ($j=i+1$, $i\geq 0$) are defined\footnote{The motivation for the superscript ``c'' (for ``classical'') shall be clarified in the Conclusion section.} by:

\begin{equation}
\label{next-to-eikonal}\left. K_i(x)\right\vert_{\rm LL}=L_x^i[p_{qq}(x)\ k_{j  i}+h_{j i}+(1-x)f_{j  i}^c+(1-x)^2g_{j  i}+{\cal O}((1-x)^3)]\ . \end{equation}
Eq.(\ref{new-kernel-full}) yields the corresponding $f_{j  i}^c(\lambda)$ coefficients in $K_i(x,\lambda)$:

\begin{eqnarray}\label{fnew}
f_{21}^c(\lambda)&=&f_{21}^c+\lambda\frac{1}{2}k_{10}^2\\
f_{32}^c(\lambda)&=&f_{32}^c-\lambda(-\frac{3}{4}k_{10}k_{21}+k_{10}h_{21})+\lambda^2\frac{1}{2}k_{10}^3\nonumber\\
f_{43}^c(\lambda)&=&f_{43}^c-\lambda\Big(-\frac{2}{3}k_{10}k_{32}+\frac{1}{2}(h_{21}-\frac{1}{2}k_{21})k_{21}+k_{10}h_{32}\Big)+\lambda^2 k_{10}^2 k_{21}\nonumber\ ,\end{eqnarray}
where I used that $h_{10}=0$, and  one should note the presence of contributions {\em quadratic} in $\lambda$.
Assuming the  $f_{j  i}^c(\lambda)$'s  vanish for $\lambda=1$, one thus derives  the relations (with $f_{10}^c=0$):

\begin{eqnarray}\label{f}
f_{21}^c&=&-\frac{1}{2}k_{10}^2=-8C_F^2\\
f_{32}^c&=&-\frac{3}{4}k_{10}k_{21}+k_{10}h_{21}-\frac{1}{2}k_{10}^3=12 C_F^2 \beta_0+32 C_F^3\nonumber\\
f_{43}^c&=&-\frac{2}{3}k_{10}k_{32}+\frac{1}{2}(h_{21}-\frac{1}{2}k_{21})k_{21}+k_{10}h_{32}-k_{10}^2 k_{21}=-\frac{44}{3}C_F^2 \beta_0^2 -64 C_F^3 \beta_0\nonumber\ ,\end{eqnarray}
which are  seen to be correct using eq.(3.26) in \cite{Moch:2009hr}. The latter equation also makes it likely that similar leading logarithmic predictions can be obtained to any order in $(1-x)$, using the {\em same} prefactor 
$p_{qq}(x)$ as in eq.(\ref{next-to-eikonal}) to define the leading term in the eikonal expansion. Indeed, 
one derives for instance \cite{GG} the ${\cal O}((1-x))^2$ coefficients in eq.(\ref{next-to-eikonal}) (with $g_{10}=0$): 

\begin{eqnarray}\label{g}
g_{21}&=&\frac{1}{3}k_{10}^2=\frac{16}{3}C_F^2\\
g_{32}&=&\frac{1}{6}k_{10}k_{21}+(f_{21}^c+\frac{1}{2}h_{21}+\frac{1}{3}k_{21})k_{10}=\frac{1}{2}k_{10}k_{21}=-8 C_F^2 \beta_0\nonumber\ ,\end{eqnarray}
which are correct  \cite{Moch:2009hr}. I note that $f_{21}^c$ and $g_{21}$ coincide (like $h_{21}$) with the splitting functions contributions.

\section{Fragmentation functions in $e^{+}e^{-}$ annihilation}

Similar results hold for physical evolution kernels associated to fragmentation functions in semi-inclusive $e^{+}e^{-}$ annihilation (SIA), provided one sets $\lambda=-1$ in the analogue of eq.(\ref{F_2_new-evolution}):

\begin{equation}
\label{Frag-new-evolution}
\frac{\partial {\cal C}_T(x,Q^2,\mu^2_F)}{\partial \ln Q^2}\,=\,\int_x^1 \frac{dz}{z}\,K_T(z,a_s(Q^2),\lambda)\,{\cal C}_T(x/z,Q^2/z^{\lambda},\mu^2_F)\ ,
\end{equation}
where ${\cal C}_T$ denotes a generic non-singlet SIA coefficient function.
I first note that threshold resummation in this case \cite{Cacciari:2001cw} leads at the leading eikonal level to an equation similar to eq.(\ref{K_x}):

\begin{equation}
\label{K_x-SIA}
K_T(x,a_s(Q^2))\sim \frac{{\cal J}\left((1-x)Q^2\right)}{1-x}\,+B_{\delta}^{SIA}(a_s(Q^2))\,\delta(1-x)\ , \end{equation}
where $x$ should now be identified to Feynman-$x$ rather then Bjorken-$x$, and I used the results of \cite{Moch:2009my} which imply that the ``physical Sudakov anomalous dimension'' ${\cal J}(Q^2)$ is  {\em the same} for structure and fragmentation functions.
The statement above eq.(\ref{Frag-new-evolution}) then follows from the following two observations: 

\noindent i) The predictions in eq.(\ref{next-to-eik-LL-1-bis}), (\ref{prediction1}), (\ref{prediction2}) and (\ref{prediction3}) depend only upon coefficients of  leading {\em eikonal} logarithms in the physical evolution kernels.

\noindent ii) Eq.(3.26) in \cite{Moch:2009hr} shows that the latter coefficients are  {\em identical} for deep-inelastic structure functions and for $e^{+}e^{-}$ fragmentation functions (consistently with the remark below eq.(\ref{K_x-SIA})), but that the coefficients of the leading {\em next-to-eikonal} logarithms are equal only up to a sign change (in an expansion in $1/r$) between deep-inelastic structure functions and fragmentation functions.

\noindent One deduces the resummation formula ($j=i+1$):
\begin{equation}
\label{all-order-LL-SIA}
\sum_{i=0}^\infty h_{j  i}^{SIA} L_x^i a_s^{i+1}=-\frac{A_1}{\beta_0}\frac{A_1 a_s}{1+a_s \beta_0 L_x}\ln(1+a_s \beta_0 L_x)\ .
 \end{equation}

\section{Conclusion}
A modified\footnote{Evolution equations involving similar kinematical rescaling factors have been suggested in the past: see e.g. eq.(3.2) in \cite{Amati:1980ch}.} evolution equation for DIS non-singlet structure functions,  analoguous to
 the one used in \cite{Dokshitzer:2005bf} for parton distributions, but which deals with the {\em physical} scaling violation and coefficient functions, has been proposed.
 It allows to relate the leading next-to-eikonal logarithmic contributions in the momentum space  physical evolution kernel to coefficients of leading eikonal logarithms at lower loop order (depending only upon the one-loop cusp anomalous dimension $A_1$), which represents the first step towards threshold resummation at the next-to-eikonal level. This result also explains the observed \cite{Grunberg:2009yi,Moch:2009hr} {\em universality} of the leading next-to-eikonal logarithmic contributions to the physical kernels of the various non-singlet structure functions, linking them to the known \cite{Moch:2008fj} universality of the eikonal contributions.
Similar results hold at the next-to-next-to-eikonal level with a proper definition of the leading eikonal piece, and can presumably be extended to leading logarithmic contributions at all orders in $(1-x)$.
Analogous results are obtained for fragmentation functions in semi-inclusive $e^{+}e^{-}$ annihilation. 

 One may ask to what extent the success of the present approach may be attributed, as suggested in \cite{Dokshitzer:2005bf,Dokshitzer:2006nm} for  the splitting functions case, to the {\em classical nature} \cite{Low:1958sn} of soft radiation. In fact, the main result of this paper for the (modified) DIS physical evolution kernel can be summarized (barring the $\delta$-function contribution) by the following equation:
  
 \begin{equation}
\label{effective-oneloop}K(x,a_s,\lambda=1)\sim \left[\frac{x}{1-x}+\frac{1}{2}(1-x)\right]{\cal J}\left((1-x)Q^2\right)\,+subleading\, logarithms\ ,
\end{equation}
where the second term (the ``subleading logarithms'') is contributed by all powers in $(1-x)$ except the leading eikonal one. The first term in eq.(\ref{effective-oneloop}) accounts for the leading logarithmic contributions  to the modified kernel (together with some subleading logarithms) to {\em all} powers in $(1-x)$ at any given loop order, and implies  leading logarithmic contributions are actually absent beyond ${\cal O}(1-x)$ power. This term has the remarkable  effective one-loop splitting function form $4C_F\, a_{phys}\left((1-x)Q^2)\right)p_{qq}(x)$, with the  ``physical coupling''  $a_{phys}(Q^2)\equiv\frac{1}{4C_F}{\cal J}(Q^2)$.

As pointed out in \cite{Dokshitzer:2006nm}, the $\frac{x}{1-x}$ part of the one-loop prefactor (eq.(\ref{prefactor}))  should be interpreted as corresponding to universal classical  radiation, a QCD manifestation  of the Low-Burnett-Kroll theorem \cite{Low:1958sn}, while the $1-x$ part represents a genuine quantum contribution. Now, it is clear that at the next-to-eikonal level, the $1-x$ part of the prefactor is irrelevant: only the ``classical'' $1/r$ part is required to  separate those leading logarithms in the standard ($\lambda=0$) physical evolution kernel  which are correctly  predicted in the present approach (the $h_{j  i}$ in eq.(\ref{next-to-eikonal})), hence ``inherited'' in the sense of  \cite{Dokshitzer:2006nm}, from the ``primordial'' ones (those which at each loop order carry the {\em same} color factors as the leading ${\cal O}(1/(1-x))$ eikonal  logarithms, and can thus be absorbed into the definition of the leading term). 
However, it appears from the results of section 4 that,  at next-to-next-to-eikonal level, the full one-loop prefactor has to be used into the definition of the leading term to properly isolate the ``inherited''  next-to-next-to-eikonal logarithms (the $f_{j  i}^c$ in eq.(\ref{next-to-eikonal})). Moreover, although the ``inherited'' $f_{j  i}^c$ are purely ``classical''  (like the $h_{j  i}$), the ``inherited'' $g_{j  i}$ at the ${\cal O}((1-x)^2)$ level are a mixture of ``quantum'' and ``classical''. Indeed, setting $f_{1 0}^q=\frac{1}{2}k_{10}$ and $f_{21}^q=\frac{1}{2}k_{21}$  (the ``quantum parts'' of the  ${\cal O}(1-x)$ coefficients), one finds
$g_{21}=g_{21}^q+g_{21}^c$, with $g_{21}^q=k_{10} f_{10}^q=\frac{1}{2}k_{10}^2$ and $g_{21}^c=-\frac{1}{6}k_{10}^2$;  and 
$g_{32}=g_{32}^q+g_{32}^c$,
with 
$g_{32}^q=\frac{1}{2}k_{21} f_{10}^q+k_{10} f_{21}^q=\frac{3}{4}k_{21}k_{10}$ and
$g_{32}^c=-\frac{1}{4}k_{21}k_{10}$. In both cases   $g_{j  i}^c=-\frac{1}{3}g_{j  i}^q$, which shows the ``inherited'' $g_{j  i}$ coefficients are actually dominantly ``quantum''.

 It can be further checked \cite{GG} that the very same first term in eq.(\ref{effective-oneloop}) also accounts for the leading logarithmic contributions to the $\lambda=-1$ modified SIA physical  evolution  kernel to {\em all} powers in $(1-x)$, which implies that the
 {\em leading logarithmic parts} of the {\em modified} DIS and SIA physical evolution  kernels satisfy Gribov-Lipatov relation \cite{Gribov:1972ri}, namely we have:

\begin{equation}
\label{GLR}\left. K(x,a_s,\lambda=1)\right\vert_{\rm LL}=\left. K_T(x,a_s,\lambda=-1)\right\vert_{\rm LL}
=p_{qq}(x)\left.{\cal J}\left((1-x)Q^2\right)\right\vert_{\rm LL}
\ , \end{equation}
where 
$\left.{\cal J}\left((1-x)Q^2\right)\right\vert_{\rm LL}= \left.A\left(a_s((1-x)Q^2)\right)\right\vert_{\rm LL}=\frac{A_1 a_s(Q^2)}{1+a_s(Q^2)\beta_0L_x}$
is the leading logarithmic contribution to eq.(\ref{j-expand}).
Indeed, once transformed back to the standard ($\lambda=0$) physical kernels,  eq.(\ref{GLR}) is consistent with eq.(3.26) in  \cite{Moch:2009hr} at least to next-to-next-to-eikonal order, and is probably correct to all orders in $(1-x)$ (with identically vanishing contributions beyond ${\cal O}(1-x)$ order).
On the other hand,  contrary to the splitting functions case where it has been checked up to three loops \cite{Mitov:2006ic,Dokshitzer:2006nm}, a full Gribov-Lipatov relation $K(x,a_s,\lambda=1)=K_T(x,a_s,\lambda=-1)$ does not seem to hold for subleading logarithms beyond  the leading eikonal level.

 The resummation of the subleading logarithmic contributions at next-to-eikonal order in eq.(\ref{effective-oneloop}), not adressed here,  remains an open  issue: the present method does not work for them, except in the conformal limit, where one recovers the results of \cite{Dokshitzer:2005bf}. 
\vspace{0.5cm}

\noindent\textbf{Acknowledgements}
I thank G. Marchesini for stimulating discussions and hospitality at the University of Milano-Bicocca where part of this paper has been written. I also wish to thank  the referees for constructive suggestions.

\end{document}